\documentclass[journal=jacsat,manuscript=article]{achemso}
\usepackage[utf8]{inputenc}
\usepackage[T1]{fontenc}

\usepackage[version=3]{mhchem} 
\usepackage{pdfcomment}
\usepackage{amssymb}
\usepackage{xcolor}
\usepackage{comment}
\usepackage[normalem]{ulem}

\newcommand{\red}[1]{\textcolor{black}{#1}}

\hypersetup{
    colorlinks=true,
    linkcolor=blue,
    citecolor=blue,
    urlcolor=blue
}

\author{Elena Sendarrubias Arias-Camisón}
\affiliation[UAM-DptoFMC]{Departamento de Física de la Materia Condensada, Universidad Autónoma de Madrid, 28049 Madrid, Spain}

\author{Maksim Lednev}
\affiliation[UAM-DptoFTMC]{Departamento de Física Teórica de la Materia Condensada, Universidad Autónoma de Madrid, 28049 Madrid, Spain}

\author{Jorge Cuadra}
\affiliation[UAM-DptoFMC]{Departamento de Física de la Materia Condensada, Universidad Autónoma de Madrid, 28049 Madrid, Spain}

\author{Raúl Gago}
\affiliation[ICMM-CSIC]{Instituto de Ciencia de Materiales de Madrid (ICMM-CSIC), 28049 Madrid, Spain}

\author{Luis Vi\~na}
\affiliation[UAM-DptoFM]{Departamento de Física de Materiales, Universidad Autónoma de Madrid, 28049 Madrid, Spain}
\alsoaffiliation[IFIMAC]{Condensed Matter Physics Center (IFIMAC), Universidad Autónoma de Madrid, 28049 Madrid, Spain}
\alsoaffiliation[INC]{Instituto Nicolás Cabrera, Universidad Autónoma de Madrid, 28049 Madrid, Spain}

\author{Francisco José García Vidal}
\affiliation[UAM-DptoFTMC]{Departamento de Física Teórica de la Materia Condensada, Universidad Autónoma de Madrid, 28049 Madrid, Spain}
\alsoaffiliation[IFIMAC]{Condensed Matter Physics Center (IFIMAC), Universidad Autónoma de Madrid, 28049 Madrid, Spain}
\alsoaffiliation[INC]{Instituto Nicolás Cabrera, Universidad Autónoma de Madrid, 28049 Madrid, Spain}

\author{Johannes Feist}
\affiliation[UAM-DptoFTMC]{Departamento de Física Teórica de la Materia Condensada, Universidad Autónoma de Madrid, 28049 Madrid, Spain}
\alsoaffiliation[IFIMAC]{Condensed Matter Physics Center (IFIMAC), Universidad Autónoma de Madrid, 28049 Madrid, Spain}
\alsoaffiliation[INC]{Instituto Nicolás Cabrera, Universidad Autónoma de Madrid, 28049 Madrid, Spain}

\author{Ferry Prins}
\email{ferry.prins@uam.es}
\affiliation[UAM-DptoFMC]{Departamento de Física de la Materia Condensada, Universidad Autónoma de Madrid, 28049 Madrid, Spain}
\alsoaffiliation[IFIMAC]{Condensed Matter Physics Center (IFIMAC), Universidad Autónoma de Madrid, 28049 Madrid, Spain}
\alsoaffiliation[INC]{Instituto Nicolás Cabrera, Universidad Autónoma de Madrid, 28049 Madrid, Spain}

\author{Carlos Antón Solanas}
\email{carlos.anton@uam.es}
\affiliation[UAM-DptoFM]{Departamento de Física de Materiales, Universidad Autónoma de Madrid, 28049 Madrid, Spain}
\alsoaffiliation[IFIMAC]{Condensed Matter Physics Center (IFIMAC), Universidad Autónoma de Madrid, 28049 Madrid, Spain}
\alsoaffiliation[INC]{Instituto Nicolás Cabrera, Universidad Autónoma de Madrid, 28049 Madrid, Spain}

\title[An \textsf{achemso} demo]
  {Spin injection of exciton-polaritons with halide perovskites at room temperature}

\begin{document}

\begin{abstract}
Exciton–polaritons, hybrid photon-exciton quasiparticles, constitute a useful platform for the study of light–matter interaction and nonlinear photonic applications. In this work, we realize a monolithic Tamm-plasmon microcavity embedding a thin film of two-dimensional halide perovskites with a tunable polymer spacer that controls the exciton–photon detuning. Angle-resolved optical spectroscopy at room temperature reveals the lower polariton branch dispersions in the linear regime for several detunings. Under circularly polarized, quasi-resonant laser excitation, the spin injection of excitons and their relaxation towards the lower polariton branch demonstrates its preservation. \red{The observed spin-polarized emission is consistent with the fast decay of polaritons: spin relaxation mechanisms become inefficient during this short time.} Our results provide promising insights into the spin control of polaritonic devices, including chiral lasers and switches. 
\end{abstract}

\section{Introduction}

Strong coupling between light and matter gives rise to exciton–polaritons, hybrid quasi-particles with both photonic and excitonic character that behave as a quantum fluid of light with properties such as lasing and non-linear optical phenomena.\cite{RevModPhys.85.299} Due to their large exciton binding energies, high oscillator strengths, and broadly tunable bandgap (from the UV to the near-infrared), lead-halide perovskite semiconductors represent an outstanding platform for polaritonics at room temperature (RT) \cite{su2021a}. Their processability and diverse structural forms, from 3D bulk crystals to low-dimensional nanoplatelets\cite{su_room-temperature_2017} and nanowires\cite{du_strong_2018}, enable strong light–matter coupling in a variety of architectures including high-quality microcavities \cite{peng_room-temperature_2022,bujalance_strong_2024,zou_continuous-wave_2025}, metasurfaces\cite{dang_tailoring_2020,liu_polariton_2023,dang_long-range_2024}, and plasmonic nanostructures\cite{symonds_emission_2007,niu_image_2015,shang_surface_2018,lu_engineering_2020,park_polariton_2022}. Recent demonstrations have reported RT lasing\cite{su_room-temperature_2017,shi_ten_2025}, long-range propagation\cite{feng_all-optical_2021,chen_unraveling_2023,kedziora_predesigned_2024,nytko_guided_2025} and engineered polariton lattices for the study of topological phenomena \cite{su_optical_2021,tao_halide_2022,wu_higher-order_2023,jin_perovskite_2024,peng_topological_2024,jin_exciton_2025}. These advances position perovskites as a promising platform for both fundamental studies of light-matter interaction as well as polariton-based applications.


 In the context of polaritonic information processing, an appealing strategy is the use of the spin degree of freedom (here, spin refers to the circular polarization degree of freedom of polaritons). Exciton–polaritons inherit their spin from both photons and excitons, enabling interesting optical phenomena such as the optical spin Hall effect \cite{leyder_observation_2007,kammann_nonlinear_2012,Gao2015,shi_coherent_2025}, Rashba–Dresselhaus spin-orbit coupling induced by the anisotropy of the optical modes\cite{rechcinska_engineering_2019,li_manipulating_2022} or that of the crystalline structure of the active material
\cite{spencer_spin-orbitcoupled_2021,ren_realization_2022,mavrotsoupakis_unveiling_2025,wang_polarization_2025}. Following this direction, RT perovskite polaritons open the way towards ultrafast spin-controlled on-chip devices\cite{Wang2026}, such as lasers, switches, and gates, while also providing synthetic gauge fields and topological photonics\cite{su_optical_2021}. A spin-dependent propagation of perovskite polaritons in RT microcavities can be implemented via resonant laser excitation, as has been demonstrated in the optical spin Hall effect in the standard (TE-TM splitting)\cite{shi_coherent_2025} and Rashba–Dresselhaus regimes \cite{liang_polariton_2024}. Resonant pump-probe spin control of polaritons has also been studied in silver-based Tamm-plasmon cavities at RT via polarization-resolved transient absorption measurements\cite{liu_dynamics_2022}.

One pending challenge for RT perovskite polaritons is to demonstrate spin-control via quasi-resonant laser driving with the exciton reservoir. The effect has been previously reported under strict non-resonant driving in other materials at liquid-helium temperatures such as II-VI and III-V semiconductors with CdTe \cite{martin_polarization_2002,Martn2004} and GaAs \cite{roumpos_signature_2009,Amo2010,PhysRevLett.107.146402,ohadi_spontaneous_2012,PhysRevB.91.075305,askitopoulos_nonresonant_2016, pickup_optical_2018, klaas_nonresonant_2019} microcavities, respectively, and at RT in organic semiconductor microcavities \cite{liang_circularly_2023,wang_polarization_2025}. Quasi-resonant spin injection at RT has been achieved in microcavity polaritons based on two-dimensional transition metal dichalcogenide excitons \cite{sun_optical_2017}. Electrical generation of spin-polarized polaritons in a perovskite metasurface has been recently demonstrated at liquid-nitrogen temperatures \cite{wang_electrically_2025}, although such a device architecture presents a more complex nanofabrication process and the electrical driving is essentially limited to continuous driving regime.

\begin{figure}[H]
    \centering
    \includegraphics[width=1\textwidth]{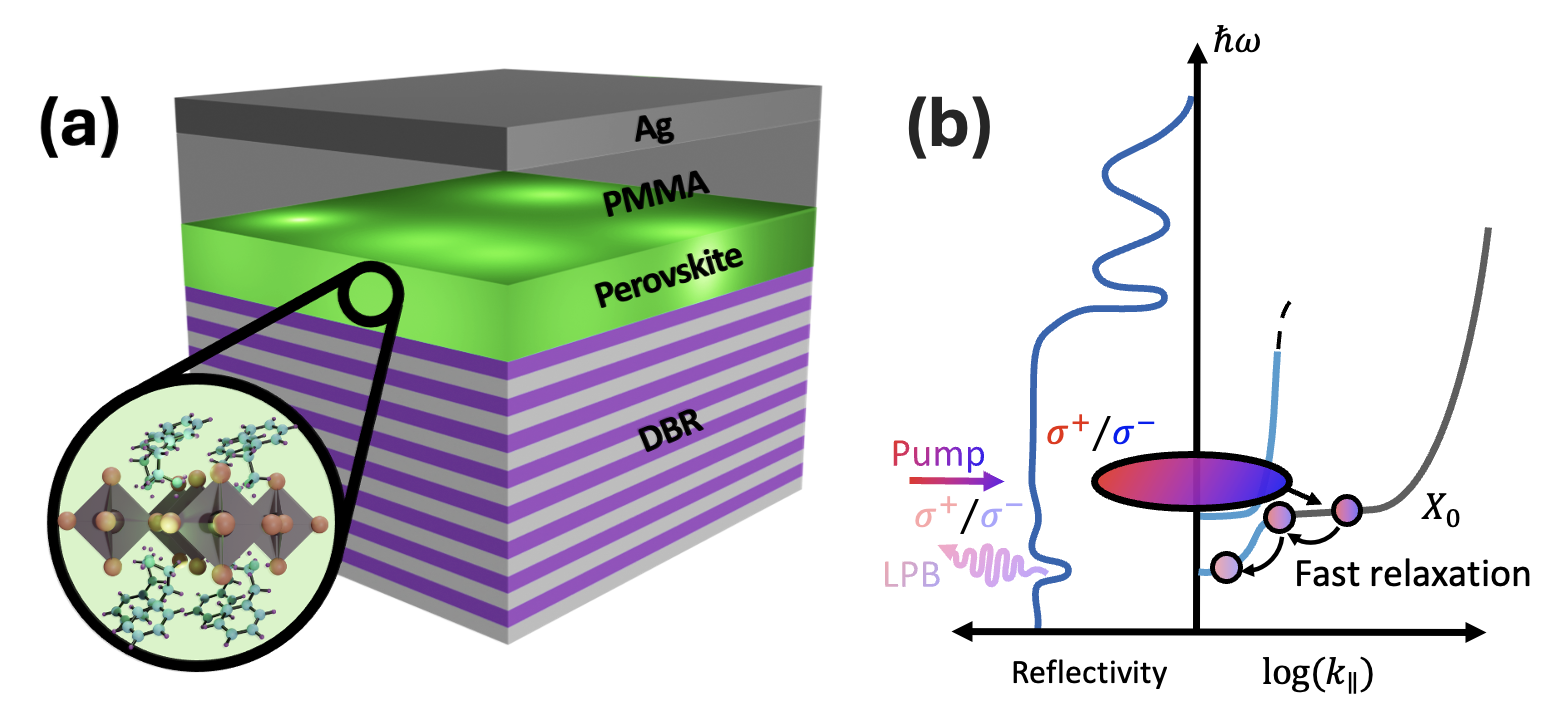} 
    \caption{\textbf{Sample design and scheme of excitation.} (a) Sketch of the sample consisting of a monolithic Tamm-plasmon structure (a thin silver layer of around 35~nm, and a DBR consisting of 8 SiO$_2$/TiO$_2$ Bragg pairs), the spacer is composed of PMMA (with tunable thickness between 45-105~nm) and a 50~nm thick layer of (PEA)$_2$PbI$_4$. (b) Scheme of the quasi-resonant spin injection used in this work. The left side profile is the microcavity reflectivity plotted against energy (vertical axis). The right part of the panel presents the polariton dispersion relation (light blue UPB and LPB dispersion relations), and the exciton dispersion (gray color) at larger momentum wavevectors (represented in log-scale).}
    \label{fig:sample_bare_modes}
\end{figure}

In this work, we engineer a Tamm-plasmon photonic microstructure in the strong coupling regime at RT\cite{lundt_room-temperature_2016,polimeno_observation_2020,shang_spinorbit_2024,gomez-dominguezMultipleEmissionPeaks2025}, and we demonstrate the spin control of the polariton emission in the linear regime. We embed a thin film of 2D phenethylammonium lead iodide ((PEA)$_2$PbI$_4$) of 50~nm thickness (corresponding to approximately 30 layers of (PEA)$_2$PbI$_4$)\cite{duTwoDimensionalLeadIIHalideBased2017} in a compact Tamm-plasmon resonator composed of a dielectric distributed Bragg reflector (DBR), a polymer (PMMA) spacing layer of varying thickness, and a 35~nm thin silver cap (see scheme of the device in Fig.~\ref{fig:sample_bare_modes}~(a)). We measure the dispersion relation of the lower polariton branch (LPB) by angle-resolved reflection and photoluminescence (PL) measurements; we also implement the full polarization tomography of the PL emission. All our experiments are performed at RT and in the linear regime (well bellow any polariton or photonic lasing threshold, see SI for further details). 

Our experimental findings show that the circularly-polarized laser pump resonantly injects excitons that, upon fast relaxation, partially preserve the spin in the LPB. We schematically represent such excitation and decay scheme in Fig.~\ref{fig:sample_bare_modes}~(b). The photonic fraction of polaritons results in a very short lifetime ($\tau_p$~$\sim$~30~-~100~fs), approaching or even outpacing the spin-flip time during the energy relaxation. This spin-flip lifetime at room temperature (RT) lies in the picosecond scale \cite{giovanni_coherent_2018}. \red{Therefore, a significant fraction of polaritons is expected to decay radiatively before the spin population becomes fully thermalized, enabling the observation of a non-zero circular degree of polarization} (Fig.~\ref{fig:sample_bare_modes}~(b)). This behavior is consistently observed across different cavity detunings (ranging from negative to positive ones) and under $\sigma^\pm$ laser excitation. 

\section{Results}

The excitonic emission spectrum shows a Lorentzian profile with a peak centered at 2.38~eV and a full-width-at-half-maximum (FWHM) of 0.08~eV (see Supplementary information Sec.~\hyperref[sec: S6]{S6}). The typical exciton lifetime of this perovskite is $\tau_X$~=~400~-~600~ps \cite{kitazawaTemperaturedependentTimeresolvedPhotoluminescence2012}. Tamm-plasmon resonators with (PEA)$_2$PbI$_4$ embedded have yielded quality factors from Q~$\sim$~70 \cite{gomez-dominguezMultipleEmissionPeaks2025} up to higher than Q~$\sim$~1000 \cite{polimeno_observation_2020}, corresponding to a cavity photon lifetime between $\tau_C$~$\sim$~20~fs \cite{gomez-dominguezMultipleEmissionPeaks2025} and $\sim$~1250~fs~\cite{polimeno_observation_2020}. Other Tamm-plasmon architectures with transition metal dichalcogenide monolayers like WSe$_2$\cite{lundt_room-temperature_2016}, hBN/Mo$S_2$ bilayer/hBN \cite{genco_femtosecond_2025} or perovskites like CsPbBr$_3$\cite{shang_spinorbit_2024} reported Q and $\tau_C$ within that range. Our reported quality factor falls within the range of reported values\cite{gomez-dominguezMultipleEmissionPeaks2025,lundt_room-temperature_2016,genco_femtosecond_2025}, with a  Q~$\sim$~100 (a cavity photon lifetime $\tau_C$~$\sim$~30~fs). From $\tau_X$ and $\tau_C$ values, and considering the accessible exciton-photon detunings of our devices, the lower polariton lifetime at $k_\parallel$~$\sim$~0 spans around 30~-~100~fs.

\begin{figure}[H]
    \centering
    \includegraphics[width=1.0\textwidth]{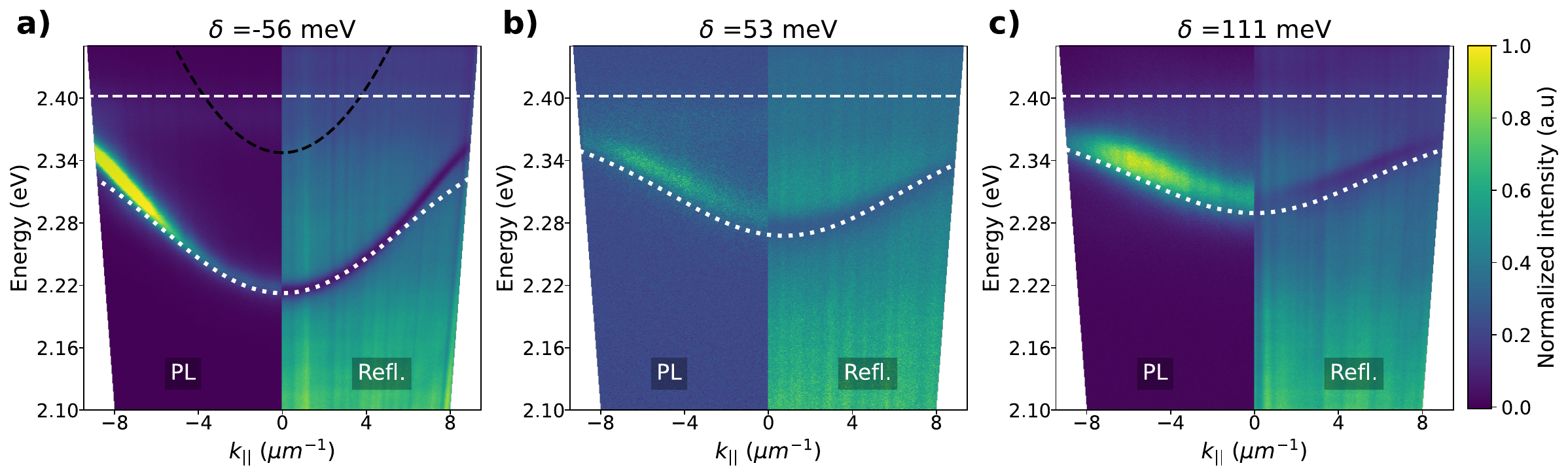} 
    \caption{\textbf{Normalised PL emission and reflectivity of the LPB dispersion relation for different exciton-photon detunings.} (a-c) Each  panel shows the normalised PL emission (left side) under non-resonant driving and the corresponding reflectivity (right side), for different detunings (indicated on the top part of each panel). The Rabi splitting is 322~$\pm$~13~meV. The fitted bare cavity and exciton modes are indicated with dashed lines, the polariton modes (dotted lines) are fittings to the dispersion relation reflectivity.}
    \label{fig:disp_rel_pol}
\end{figure}

We study the dispersion relation of polaritons under white light excitation, retrieving the reflectivity map of the device (see right panels in Figs.~\ref{fig:disp_rel_pol}~(a-c) for several detunings). We also implement PL experiments under non-resonant (2.76~eV), continuous-wave laser excitation and an excitation power density of 8.86~$\mu$W/$\mu$m$^2$ (left panels in Figs.~\ref{fig:disp_rel_pol}~(a-c)). In this excitation scheme, the sample is operated in the linear regime (well below a potential lasing threshold, see the Supp. information for further details). The three exciton-photon detunings in these samples are -56~$\pm$~10~meV, 53~$\pm$~15~meV, and 111~$\pm$~18~meV, corresponding to 40, 85 and 105~nm PMMA thicknesses, respectively.

Following a standard two-mode coupled polariton model (see Suppl. information Sec.~\hyperref[sec: S2]{S2}), we retrieve a vacuum Rabi splitting of 322~$\pm$~13~meV, which is similar to other Tamm-plasmon resonator architectures recently reported with perovskite excitons \cite{polimeno_observation_2020, shang_spinorbit_2024, gomez-dominguezMultipleEmissionPeaks2025}. Similar cavity architectures with this type of perovskite show a Rabi splitting of 100~meV at cryogenic temperatures\cite{polimeno_observation_2020,gomez-dominguezMultipleEmissionPeaks2025}. We note that the upper polariton branch (UPB) is outside of the measured frequency range in the reflectivity and PL experiments of Fig.~\ref{fig:disp_rel_pol}. We obtain similar reflectivity results to this figure using the Transfer matrix method (see Suppl. information Sec.~\hyperref[sec: S3]{S3}). The exciton/photon character of the LPB is manifested in the variation of its PL FWHM at zero in-plane momentum: from panels (a-c) in Fig.~\ref{fig:disp_rel_pol}, these values are $20\pm1$, $30\pm1$, $40\pm1$ meV from negative to positive detuning.

\begin{figure}[H]
    \centering
    \includegraphics[width=0.85\textwidth]{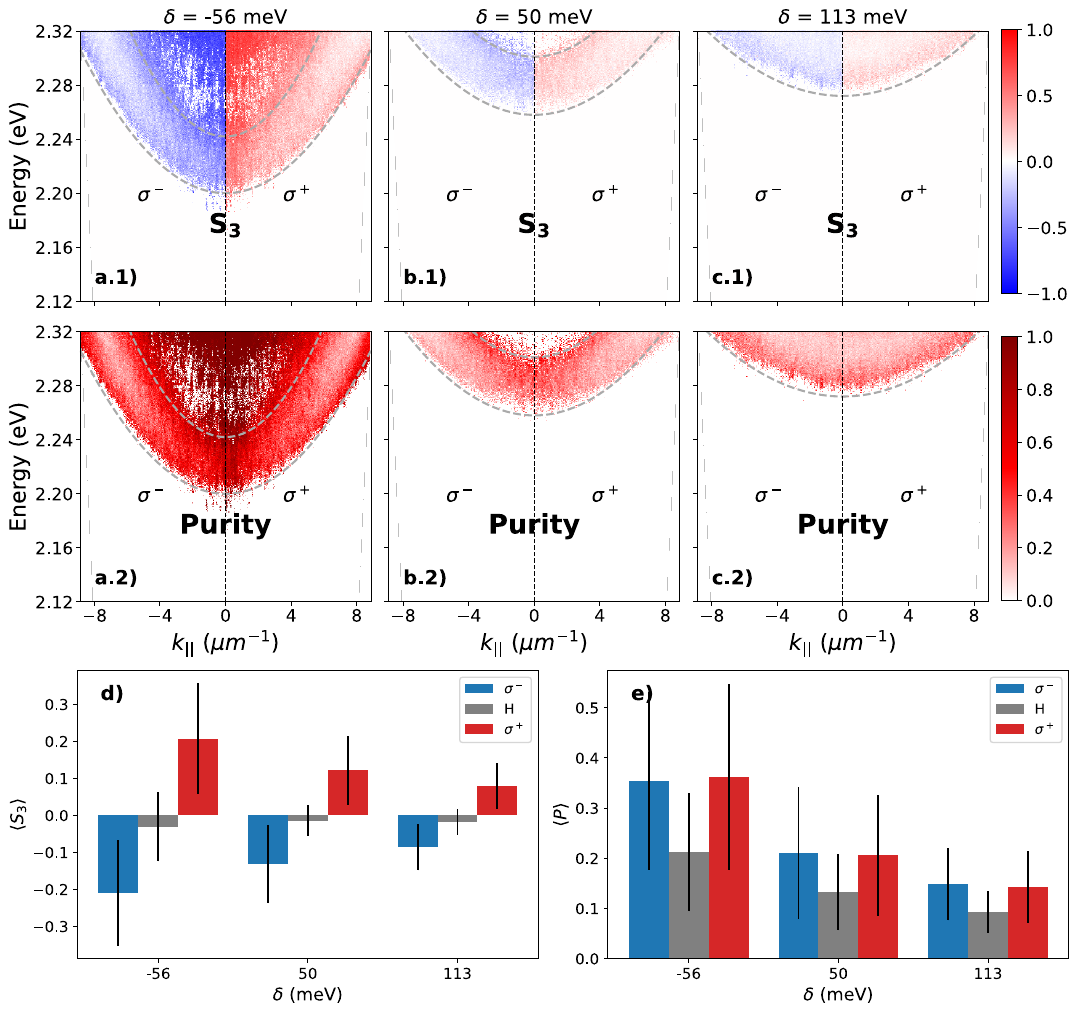} 
    \caption{\textbf{Spin injection and polarization purity of the LPB emission for different detunings}. (a1-c1) $S_3$ of the LPB dispersion relation under $\sigma^-$/$\sigma^+$ laser excitation in the left/right side of each panel, respectively. The detunings are indicated in the top part of each panel. (a2-c2) Corresponding polarization purity under the same excitation conditions. The pump power density for these measurements is 8.39~$\mu$W/$\mu$m$^2$. (d) Averaged values of $\langle S_3 \rangle$ extracted from the first row, including the case under linearly polarized pump (gray bar). \red{The averaged area is delimited by dashed gray lines in the dispersion relation.} (e) Corresponding averaged polarization purity $\langle P \rangle$. \red{The error bars represent the standard deviation of $\langle S_3 \rangle$ and $\langle P \rangle$ in the regions of interest marked in panels (a.1)-(c.1) and (a.2)-(c.2), respectively.}}
    \label{fig:disp_rel_S3_P}
\end{figure}

In order to assess the spin injection of RT polaritons, we implement polarization-resolved experiments, under quasi-resonant laser excitation with circular ($\sigma^{\pm}$) and linear ($H$) polarization. The circularly-polarized laser allows to resonantly excite excitons with specific $\pm$~1 spins, which relax towards the LPB, partially conserving the initial injected spin (see excitation and decay scheme in Fig.~\ref{fig:sample_bare_modes}~(b)).\cite{Damen1991} We implement polarization tomography experiments of the LPB emission (see further details in \hyperref[sec:Methods]{Sec. Methods} and Suppl. information Sec.~\hyperref[sec: S4]{S4}), giving special attention to the degree of circular polarization, $S_3=(I_{\sigma^+} {-} I_{\sigma^-})/(I_{\sigma^+} {+} I_{\sigma^-})$, where $I_{\sigma^\pm}$ is the collected $\sigma^{\pm}$ PL intensity. Across the different detunings, the LPB emission exhibits a co-polarization with the pump (2.38~eV). The $S_3$ maps of the LPB dispersion relation under $\sigma^\pm$ pump are shown for several detunings on the right and left side of Figs.~\ref{fig:disp_rel_S3_P}~(a1-c1), respectively. For the sake of clarity, a cutoff filter for the total intensity $I_{\sigma^+}$~+~$I_{\sigma^-}$ values smaller than 10$~\%$ of the maximum total intensity is applied to the $S_3$ dispersion relation maps. In the Suppl. information Sec.~\hyperref[sec: S4]{S4}, we also include the dispersion relation under horizontal polarization excitation, which equally populates both hot carrier spins, and consequently leaves an $S_3$ value in the LPB emission close to zero. For the sake of completeness, this supplemental section also compiles the dispersion relation for all the Stokes components.

We observe that $S_3$ remains approximately constant along the LPB dispersion. Slightly enhanced values appear in regions where the emitted intensity is weak, as a result of normalizing $S_3$ in areas where the total intensity is low. In these experiments under resonant exciton driving, the excitation laser fills the back focal plane of the objective, enabling resonant exciton pumping over the full range of optically accessible in-plane momenta. In order to isolate the LPB emission from residual scattered laser, the emitted light is spectrally dispersed and the detection window on the CCD is chosen such that the exciton resonance lies outside the recorded energy range. This procedure yields an LPB intensity map largely free from scattered laser light, at the expense of partially truncating the high-energy side of the LPB dispersion, as visible in Fig.~\ref{fig:disp_rel_pol}. We remark that the residual $|S_3|$ intensity values outside of the LPB region (delimited within the gray lines for the most negative detuning) most likely arises from residual scattered laser light. For the sake of clarity, we also include in the Supplementary the corresponding maps of $S_0$. We observe that the $|S_3|$ values along the LPB decay as polaritons become more excitonic (going towards positive detunings), starting from $|S_3|\sim20\%$ for $\delta\simeq-56$ meV, down to $|S_3|<10\%$ for $\delta\simeq50$ meV. The comparison of the $|S_3|$ magnitude with the other Stokes components $|S_{1,2}|$ shows that the degree of circular polarization is the dominating component of the emission (see Supplementary). \red{These results indicate that the short-lived (photon-like) polaritons retain the injected spin more efficiently than exciton-like polaritons at higher detunings, consistent with the shorter polariton lifetime of the former ones.}

The full polarization tomography allows us to retrieve the polarization purity of the LPB emission, calculated as $P$~=~$\sqrt{S_1^2 + S_2^2 + S_3^2}$, where $S_{1/2}$ is the linear/diagonal degree of polarization, respectively. These purity maps of the LPB dispersion relation are shown in Figs.~\ref{fig:disp_rel_S3_P}~(a2-c2), displaying a progressive depolarization of the emission when the LPB becomes more excitonic. These results show that, under circularly polarized driving, excitons are resonantly injected with $|S_3|\sim1$ and the fast relaxation process towards the LPB is dominated by a strong depolarization.

The decrease of $|S_3|$ and general depolarization of the emission as a function of increasing detunings is compiled in the following Figs.~\ref{fig:disp_rel_S3_P}(d,e). From the maps of $S_3$ and $P$, we calculate the corresponding average values of the LPB emission for each detuning and $\sigma^{+}$, $H$ and $\sigma^{-}$ pumping, see Figs.~\ref{fig:disp_rel_S3_P}(d,e), respectively. In these panels, the $\langle \cdot \rangle$ symbol stands for the average value of $S_3$ and $P$ in the area delimited by the dashed gray lines in panels (a.1-c.1) and (a.2-c.2). \red{The error bars correspond to the standard deviation of the $S_3$ and polarization purity $P$ values within the selected regions of interest of the LPB dispersion relation, delimited by the dashed gray lines in Figs.~\ref{fig:disp_rel_S3_P}(a1-c1) and ~\ref{fig:disp_rel_S3_P}(a2-c2), respectively. They therefore quantify the spread of the polarization values within the selected momentum-energy region. We note that the three detunings shown in Fig. ~\ref{fig:disp_rel_S3_P} correspond to three independently fabricated microcavity devices with different PMMA spacer thicknesses. The same qualitative behavior was consistently reproduced across several tens of fabricated devices spanning a large range of exciton-photon detunings.} We remark that the area above the LPB in the most negative detuning (Figs.~\ref{fig:disp_rel_S3_P}(a.1-2)) is excluded from the analysis as the data points do not represent meaningful physical values. In the Suppl. information Sec.~\hyperref[sec: S6]{S6} the polarization analysis of the bare exciton emission is shown: no spin conservation is observed for the bare perovskite exciton under $\sigma^{\pm}$ driving. \red{A significant enhancement of the $S_3$ values under quasi-resonant exciton excitation is observed compared to strictly non-resonant excitation. We primarily attribute this enhancement to the more efficient injection of spin-polarized excitons into the reservoir.}

\begin{figure}[H]
    \centering
    \includegraphics[width=0.85\textwidth]{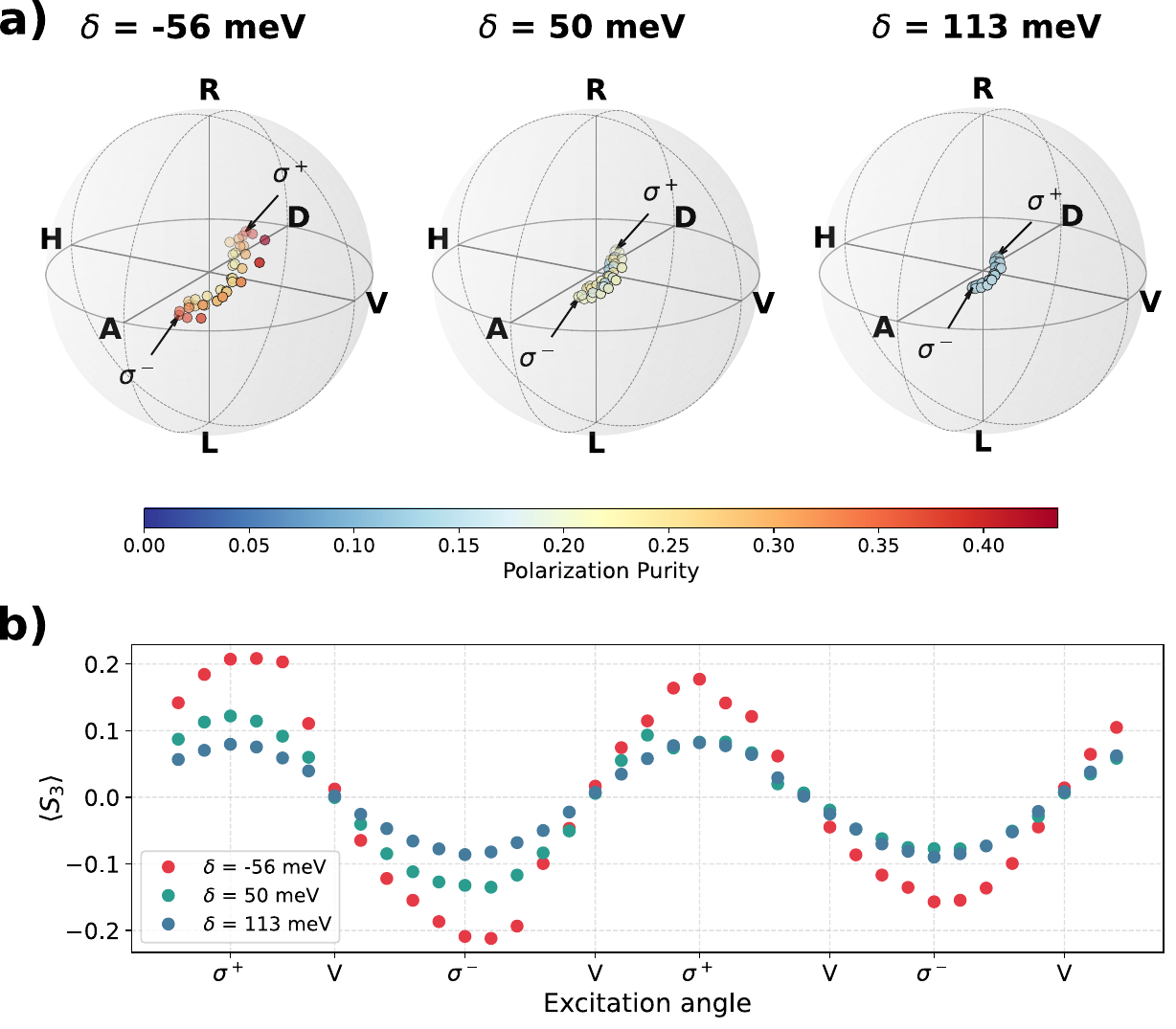} 
    \caption{\textbf{Control of the Stokes parameters of the LPB emission as a function of the excitation helicity.} 
    (a) Poincaré sphere representation of the averaged Stokes parameters for different QWP angles of the excitation, shown for the three detunings presented in Fig.~\ref{fig:disp_rel_S3_P}. The points corresponding to maximum spin injection, i.e. $\sigma^\pm$ excitation, are indicated by arrows. The color of each data point encodes the polarization purity (false-color scale below panel (a)). As the detuning becomes more positive, the polarization purity decreases, leading to polarization trajectories that contract towards the center of the sphere. (b) Averaged $S_3$ as a function of the excitation QWP angle for the three detunings.}
    \label{fig:S3_int_exc_angle}
\end{figure}

To fully characterize the spin injection, the Stokes components are measured and averaged for different pump helicities, varying angles of the excitation quarter wave plate (QWP). Fig.~\ref{fig:S3_int_exc_angle}(a) shows the polarization trajectory of the LPB emission in Poincaré sphere as a function of the pump helicity  for the three detunings shown in Fig.~\ref{fig:disp_rel_S3_P}. The data points under $\sigma^\pm$ excitation are labeled in the three trajectories of panel (a)  to identify the points of maximum spin injection. The color of the data points encodes the degree of polarization purity (see bottom false color scale), revealing that the spin injection decreases from depolarization effects as the detuning becomes more positive, as equivalently shown in Fig.~\ref{fig:disp_rel_S3_P}. For the sake of clarity, Fig.~\ref{fig:S3_int_exc_angle}(b) displays the averaged $S_3$ component as a function of the excitation angle (continuously varying along $\sigma^+$-linear polarization-$\sigma^-$): the amplitude of $S_3$ is reduced as the detuning becomes more positive, reflecting a decrease in the spin retention. The complete set of $S_3$ maps of the LPB dispersion relations for all detunings are shown in Suppl. information Sec.~\hyperref[sec: S5]{S5}.

The RT spin relaxation time of 2D phenethylammonium lead iodide perovskite excitons varies from tens of microseconds at cryogenic temperatures\cite{kirstein_coherent_2023} down to the 1 ps time scale at RT\cite{giovanni_coherent_2018}. We attribute the strong depolarization and small $S_3$ values of the LPB to the fast RT spin decoherence mechanisms. Importantly, and as pointed out before, no spin injection has been observed for the bare exciton (see Suppl. information Sec.~\hyperref[sec: S6]{S6}), as the lifetime of the exciton is orders of magnitude above the polariton lifetime, leading to the complete loss of spin coherence and, consequently, a full depolarization of the emission. \red{Our results suggest that polaritons open an additional relaxation channel for high-energy excitons that may enable the partial retention of the injected spin in the LPB.}

To substantiate our qualitative understanding, we developed a simplified model based on semiclassical rate equations to estimate the attainable value of |$S_3$| parameter under our experimental conditions (see Methods for details). The model yields
\begin{equation}
    |S_3| = \frac{1}{1+\eta}, \quad \text{with} \quad \eta = \frac{2\tau_X}{\tau_S \left[ 1 + (P - 1) \cdot \mathrm{QY} \right]},
\end{equation}

where $\tau_X$ and $\tau_S$ denote the exciton lifetime and spin relaxation time, respectively, QY is the radiative quantum yield of the bare excitons, and P represents the Purcell factor of the cavity. Using representative parameters for (PEA)$_2$PbI$_4$ ($\tau_X$~=~420~ps, $QY$~=~0.3, $\tau_S$~=~1~ps), the model indicates that a Purcell factor on the order of $P{\sim}100$ is required to reproduce the experimentally observed $S_3$ values. In planar cavities, however, the integrated Purcell factor is typically limited to only a few units~\cite{Reid1992}, consistent with our simulations (see Supplementary Information Sec.~\hyperref[sec: S3]{S3}). \red{Given the quasi-resonant nature of the excitation, we assume that the excitons participating in polariton formation largely retain their in-plane momentum $k_{\parallel}$. Although this momentum conservation is not directly verified experimentally in the present work, it provides a physically motivated framework to interpret the observed spin injection. Under this assumption, the relaxation into the polaritonic state is governed by the $k_{\parallel}$-resolved Purcell factor, in contrast to incoherent pumping, where the integrated Purcell factor dictates the dynamics.} In our system, the $k_{||}$-resolved factor reaches a value of $P = 80$ (see inset Supplementary Fig. S3(b)). Using this value, our theoretical estimate yields $|S_3| = 2.9 \times 10^{-2}$. In contrast, without the cavity ($P = 1$), this value drops to $|S_3| = 1.1 \times 10^{-3}$. Matching the experimentally observed $|S_3|$ to the model would require a higher Purcell factor, and quantum yield or a smaller ratio $\tau_X/\tau_S$. While the cavity effects of our simulations capture the spin injection variation, the simplicity of this model and the uncertainty of the parameters explain the discrepancy between theory and experiments.


In doped perovskites, spin-polarized excitation could cause magnetization reordering of dopants, from which optical spin-control could be enhanced\cite{neumann_manganese_2021}. Lowering the temperature reduces phonon population, thereby extending spin lifetimes, as reported for non-resonant excitation in II-VI CdTe \cite{martin_polarization_2002} or III-V GaAs \cite{roumpos_signature_2009,ohadi_spontaneous_2012, askitopoulos_nonresonant_2016, pickup_optical_2018, klaas_nonresonant_2019} samples. While our work focuses on RT operation, future studies could explore moderate cooling or phonon engineering to assess the spin injection with longer spin-flip time in these perovskites. For developing RT polariton devices with slower spin-flip time, we identify several metal halide perovskite excitons with relaxation times ranging on the order of~$\sim$~10~-~100~ps,\cite{haqueSpinEffectsMetal2025} being (S-MB)$_2$(MA)$_3$ Pb$_4$I$_{13}$\cite{abdelwahab_two-dimensional_2024} the perovskite kind that provides the largest RT spin relaxation time ($\sim$~70~ps).

\section{Conclusions}
\label{sec:Conclusions}

This study demonstrates RT spin injection of exciton--polaritons in a Tamm-plasmon microcavity embedding a 2D layered (PEA)$_2$PbI$_4$ perovskite. Under circularly polarized quasi-resonant excitation, polarization-resolved PL measurements show that the circular polarization of the LPB emission partially follows the helicity of the excitation laser, yielding a measurable degree of circular polarization across different exciton--cavity detunings. The effect becomes weaker for more positive detunings, where the LPB acquires a larger excitonic character, revealing stronger depolarization during the relaxation process towards the polariton states.

\red{Importantly, this spin-polarized emission is absent in the bare exciton, suggesting that the polaritonic decay channel enables partial spin retention before complete RT spin depolarization takes place. The comparison between the bare exciton and the strongly coupled system therefore supports the interpretation that the ultrafast polariton lifetime contributes to the conversion of a fraction of the injected spin polarization into circularly polarized LPB emission.}

These results establish halide perovskite Tamm-plasmon microcavities as a promising platform for RT spin-based polariton devices controlled under quasi-resonant excitation. Future experiments could explore perovskite compositions with longer spin relaxation times, as well as more advanced photonic architectures, including engineered metasurfaces\cite{wang_electrically_2025} and bilayer metasurface Fabry--Pérot cavities\cite{alagappan_fabry-perot_2024}, to enhance spin retention and further control information encoding in the polariton spin degree of freedom.

\section{Acknowledgments}
\label{sec:acknowledgments}

This work was funded by the European Union (ERC, EnVision, project number 101125962). Views and opinions expressed are however, those of the author(s) only and do not necessarily reflect those of the European Union or the European Research Council Executive Agency. Neither the European Union nor the granting authority can be held responsible for them. F.P. acknowledges funding from the Spanish AEI under grant agreements PID2022-141579OB-I00, TED2021-131018B-C21, and CNS2023-143577. In addition, we acknowledge the support from the “(MAD2D-CM)-UAM” project funded by Comunidad de Madrid, by the Recovery, Transformation and Resilience Plan, and by NextGenerationEU from the European Union. L.V and C.A-S. acknowledge the support from the projects from the Ministerio de Ciencia e Innovaci\'on PID2023-148061NB-I00 and PCI2024-153425, the project ULTRABRIGHT from the Fundaci\'on Ram\'on Areces and the Grant “Leonardo for researchers in Physics 2023” from Fundaci\'on BBVA. CA-S acknowledges the support from the Comunidad de Madrid fund “Atracci\'on de Talento, Mod. 1”, Ref. 2020-T1/IND-19785. M.L., F.J.G.V., and J.F. acknowledge support from the Ministerio de Ciencia e Innovación - Agencia Estatal de Investigación through FPI Grant PRE2021-098978 to M.L. with support from ESF+, as well as Grants PID2021-125894NB-I00, EUR2023-143478, PID2024-161142NB-I00. We acknowledge Attocube for the support with the nanopositioning system of the sample. L.V., F.J.G-V., J.F., F.P, and C.A-S acknowledge support from the “María de Maeztu” Program for Units of Excellence in R\&D (CEX2023-001316-M). We thank P. Vaquer de Nieves for his contributions during the initial stages of this work.

\section{Methods}
\label{sec:Methods}
\textbf{Sample fabrication}: The DBR mirror is synthesized using reactive magnetron sputtering and consists of 8 bilayers of TiO$_2$ and SiO$_2$ piled up sequentially in the deposition chamber with a last layer of SiO$_2$ on the surface where the perovskite layer is deposited. The growth rate for each material has been calibrated by thickness measurements in monolithic films as extracted from spectroscopic ellipsometry with a Sopra GES-5E device. In particular, each TiO$_2$ layer is of 60~($\pm5$)~nm thick. They are grown by DC reactive magnetron sputtering from a pure Ti (99.99~$\%$) target. The discharge power was 150~W, resulting in a growth rate of $\sim$~2~nm/min. Instead, every SiO$_2$ layer is of~95~($\pm5$)~nm. They are grown by pulsed DC reactive magnetron sputtering from a pure Si (99.99$\%$) target. The pulse discharge was set at 100 kHz and 40$\%$ duty cycle with an overall power of 75~W, resulting in a growth rate of $\sim$~6~nm/min. For the growth of both layer materials, the working pressure was 0.3~Pa with an Ar/O$_2$ gas mixture of 28/2 standard cubic centimeter per minute flux ratio. The (PEA)$_2$PbI$_4$ perovskite has been synthesized chemically by stoichiometric dilution in DMF and consists of single PbI$_4$ layers separated by the PEA molecules, allowing RT PL emission. The perovskite solution is deposited on the DBR using spin coating. A protective layer of PMMA, prepared as a 4$\%$ solution in anisole, is applied over the perovskite also using spin coating. The layer serves both to protect the perovskite from air and to define the position of the cavity mode. Various detunings are obtained by adjusting the thickness of the PMMA layer. In Figs.~\ref{fig:sample_bare_modes}~(a-c), the PMMA spin-coating process is carried out at speeds of 5000~rpm, 4000~rpm, and 3000~rpm, respectively, leading to 40, 80 and 105~nm thicknesses, respectively. Lastly, a 35~nm thin layer of silver is e-beam evaporated to close the resonator.

\textbf{Experimental setup:} Optical spectroscopy is performed using a custom-built optical setup (see Suppl. information Sec.~\hyperref[sec: S1]{S1} for details) featuring two main light sources to probe the cavity. The first laser is a CW laser at 2.38~eV for quasi-resonant excitation, matching the PL energy of perovskite excitons. The second is a white-light halogen lamp, used for reflectivity measurements. The quasi-resonant laser passes through a polarizer and a quarter wave plate to have the choice to continuously transform the helicity of the pump, from linear polarization to $\sigma^+$, or $\sigma^-$. All excitation sources are focused onto the silver mirror of the cavity using a 0.75 numerical aperture, 5.2~mm working-distance objective. The cavity is mounted on XYZ piezo stages, allowing precise positioning and exploration of different regions, which is essential to have access to the slight detuning variations that could be found across the sample. A momentum space lens has been installed to switch between real-space and momentum-space imaging. The emission is focused by a lens onto the entrance slit of a spectrometer, which is connected to a CCD camera, to resolve it in energy and angular or real space; this lens allows for the two-dimensional reconstruction of momentum or real space.

For the polarization tomography, we decompose the emission intensity along the six poles of the Poincaré sphere, grouped into three orthogonal bases: $H$/$V$, $D$/$A$, and $\sigma^+$/$\sigma^-$ complete the tomography polarization measurements on the emission,which reconstructs the Stokes parameters, defined as $S_0$~=~$I_{\parallel} + I_{\perp}$ and $S_i = \frac{I_{\parallel} - I_{\perp}}{S_0}$, where, for $i = 1, 2, 3$, the symbols $\parallel$ and $\perp$ correspond to the polarization pairs $H$/$V$, $D$/$A$, and $\sigma^+$/$\sigma^-$, respectively. Here, $S_0$ represents the total emission intensity, while the $S_i$ denote the three Stokes components. The emission passes through a set of $\lambda$/4, $\lambda$/2 wave-plates analyzing the emission polarization.

For the PL measurements shown in Fig.\ref{fig:disp_rel_pol}, we employed a non-resonant laser at 2.76~eV instead of the quasi-resonant laser and long-pass filters are added in the detection path to suppress residual scattered laser light before the signal enters the spectrometer and CCD camera.

Rate equations model. To theoretically describe the population dynamics of the excitonic reservoir and lower polaritons, we employed a phenomenological semiclassical rate-equation model. The coupled differential equations governing the polarization-resolved populations of the excitonic reservoir ($n_{X,\pm}$) and lower polaritons ($n_{p,\pm}$) are given by:
\begin{subequations}
    \begin{align}
        \dot{n}_{X,\pm} &= I_{\pm}
        - \frac{1}{\tau_{nr}} n_{X,\pm}
        - \frac{P}{\tau_{r}} n_{X,\pm}
        - \frac{1}{\tau_{S}} (n_{X,\pm} - n_{X,\mp}), \\
        \dot{n}_{p,\pm} &= \frac{P}{\tau_{r}} n_{X,\pm}
        - \frac{1}{\tau_p} n_{p,\pm},
    \end{align}
\end{subequations}
where $I_{\pm}$ denotes the polarization-dependent pumping term, $\tau_{nr}$ and $\tau_r$ are the nonradiative and radiative exciton lifetimes, respectively, $\tau_S$ is the exciton spin relaxation time, $\tau_p$ is the polariton lifetime, and P is the Purcell factor of the cavity.

In this model, excitons scatter into the polaritonic reservoir through a radiative pumping process occurring at a rate $P$/$\tau_r$. Owing to the ultrafast cavity decay ($\tau_c$~$\approx$~25~fs), we assume that polaritons are emitted almost instantaneously, allowing us to neglect any spin depolarization processes within the polaritonic reservoir.

The steady-state solution of the above system yields the following expression for $S_3$:
\begin{subequations}
    \begin{align}
        S_3 &\equiv \frac{n_{p,+} - n_{p,-}}{n_{p,+} + n_{p,-}} = \frac{1}{1 + \eta}, \\
        \eta &= \frac{2\tau_X}{\tau_S \left[ 1 + (P - 1) \cdot \mathrm{QY} \right]},
    \end{align}
\end{subequations}
where we set $I_+$~=~I and $I_-$~=~0. To obtain the final expression for $\eta$, we used the relations 1/$\tau_X$ = 1/$\tau_r$ + 1/$\tau_{nr}$ and $QY$~=~$\tau_{nr}$/$\tau_X$.

\bibliography{bib_spin_memory}

\end{document}